\documentclass[12pt]{iopart}

\usepackage{iopams}
\usepackage{dsfont}
\usepackage{amsthm}
\usepackage{graphicx}
\theoremstyle{definition}
\newtheorem{Def}{Definition}[section]

\theoremstyle{plain}

\newtheorem{Thm}[Def]{Theorem}

\theoremstyle{remark}

\newtheorem*{Pf}{Proof}
\newtheorem*{Thm*}{}

\newcommand{\dd}{\mathrm{d}}
\newcommand{\ee}{\mathrm{e}}
\newcommand{\ii}{\mathrm{i}}
\newcommand{\rh}{r_\mathrm{h}}

\newcommand{\rW}{\varrho}
\newcommand{\zW}{\zeta}
\newcommand{\pmN}{^\pm_\mathrm{N}}
\newcommand{\pN}{^+_\mathrm{N}}

\newcommand{\pS}{^+_\mathrm{S}}

\newcommand{\beq}{\begin{equation}}
\newcommand{\eeq}{\end{equation}}
\newcommand{\bea}{\begin{eqnarray}}
\newcommand{\eea}{\end{eqnarray}}

\begin{document}

\title[The inner Cauchy horizon of axisymmetric and stationary black holes]
 {The inner Cauchy horizon of axisymmetric and stationary black holes
 with surrounding matter\footnote{This paper is dedicated to Reinhard
 Meinel on the occasion of his 50th birthday.}}

\author{Marcus Ansorg and J\"org Hennig}

\address{Max Planck Institute for Gravitational Physics,\\
Am M\"uhlenberg 1, D-14476 Golm, Germany}
\eads{\mailto{mans@aei.mpg.de}
      and \mailto{pjh@aei.mpg.de}}

\begin{abstract}
We investigate the interior of regular axisymmetric and stationary black holes surrounded by matter and
find that for non-vanishing angular
momentum of the black hole the space time can always be extended regularly up to and including
an inner Cauchy horizon. We provide an explicit
relation for the regular metric at the inner Cauchy horizon in terms of that
at the event horizon. As a consequence, we obtain the universal equality
$(8\pi J)^2 = A^+ A^-$ where $J$ is the black hole's angular momentum
and $A^-$ and $A^+$ denote the horizon areas of inner Cauchy and event
horizon, respectively. We also find that in the limit $J \to 0$
the inner Cauchy horizon becomes singular.
%\\(Manuscript date: \today)
\end{abstract}


%%%%%%%%%%%%%%%%%%%%%%%%%%%%%%%%%%%%%%%%%%%%%%%%%%%%%%%%%%%%%%%%%%%%%%%%%
\section{Introduction}\label{Introduction}

An interesting feature of the well-known Kerr solution is the
existence of a Cauchy horizon $\mathcal H^-$ inside the black hole. 
While outside the black hole the two Killing vectors 
$\xi$ and $\eta$, describing stationarity and axisymmetry, can always be 
linearly combined to form a time-like vector, any such non-trivial
linear combination inevitably  
leads to a space-like vector when performed in some interior neighbourhood of 
the event horizon. As a consequence, the axisymmetric and stationary
Einstein equations,  
being elliptic in the black hole's exterior, become hyperbolic in its
interior.  
Hence, for the Kerr solution a boundary of the future domain of dependence 
of the event horizon $\mathcal H^+$ can be identified, and this is the
inner Cauchy horizon in question.  
As far as the mathematical form of the field equations is concerned,
$\mathcal H^-$ is completely equivalent to $\mathcal H^+$. However, from
a physical point of view, the 
inner Cauchy horizon is a {\em future} horizon whereas the event horizon
is a {\em past} one. While the  
space-time is always regular at $\mathcal H^+$, it is regular at
$\mathcal H^-$ only if the black hole's angular momentum $J$ does not
vanish, i.e. for $J\to 0$ the horizon $\mathcal H^-$ becomes singular.

In this paper we find that this picture also holds true for the black hole's interior
of general axisymmetric and stationary space-times which contain a regular black
hole and surrounding matter\footnote{Note that we concentrate here on pure gravity (i.e.~no electromagnetic fields) with vanishing cosmological constant.}. We are able to provide an
explicit relation between the metric at the inner Cauchy horizon and
that at the event horizon. Moreover a universal equality  
\beq\label{UnivEquality}
	(8\pi J)^2 = A^+ A^-
\eeq 
results where $A^-$ and $A^+$ 
denote the horizon areas of inner Cauchy and event horizon respectively. 

%We note that the calculation of the complete metric inside the black
%hole can be reduced to the solution of a linear integral equation. This
%solution procedure emerges from the fact that the axisymmetric and
%stationary Einstein equations permit a treatment in terms of soliton
%methods. For more details we refer to \cite{Hennig3} in which this technique
%is presented in detail for the Gowdy space-times which are closely
%related to the interior space-time region of an axisymmetric and stationary
%black hole. 

% We note that the interior space-time region of axisymmetric and
% stationary black holes is closely related to Gowdy space-times. For more
% details we refer to \cite{Hennig3}.

The paper is organized as follows. In Sec.~2 we recall Weyl's
coordinates which cover an exterior vacuum vicinity of the black hole. In order
to describe the black hole interior we introduce Boyer-Lindquist type
coordinates. We revisit the formulation in terms of the complex Ernst
potential for which the Einstein equations can be combined in the
complex Ernst equation. In Sec.~3 we use this formulation to write the Ernst potential $f$,
describing the exterior vicinity of a black hole, as a {\em B\"acklund transform}
of another Ernst potential $f_0$ which corresponds to a
space-time without a black hole, but with a regular central
vacuum region\footnote{A well-known example for this procedure is the construction of the Kerr solution 
from Minkowski space $f_0=1$, see e.g. \cite{Neugebauer}.}. 
In Sec. 4 we take the B\"acklund representation in
order to expand $f$  into the interior of the black hole. It turns out
that by utilizing appropriate symmetry properties of $f_0$, an explicit
formula of the Ernst potential $f$ at the inner Cauchy horizon $\mathcal H^-$ 
in terms of that at the event horizon $\mathcal H^+$ can be derived. Finally, from this formula we are able to 
conclude the universal equality (\ref{UnivEquality}), see Sec.~5.

%In this paper we use units in which the speed of light as well as
%Newton's constant of gravitation are unity. 

%%%%%%%%%%%%%%%%%%%%%%%%%%%%%%%%%%%%%%%%%%%%%%%%%%%%%%%%%%%%%%%%%%%%%%%%%
\section{Weyl coordinates and Ernst equation}\label{WeylCoords}

%As depicted above, in this paper the Ernst formulation plays a
%fundamental role as  
%it permits us to express the metric quantities of the black hole
%space-time in terms of that of  
%a regular central vacuum region. 

\begin{figure}
 \centering
 \includegraphics[scale=0.9]{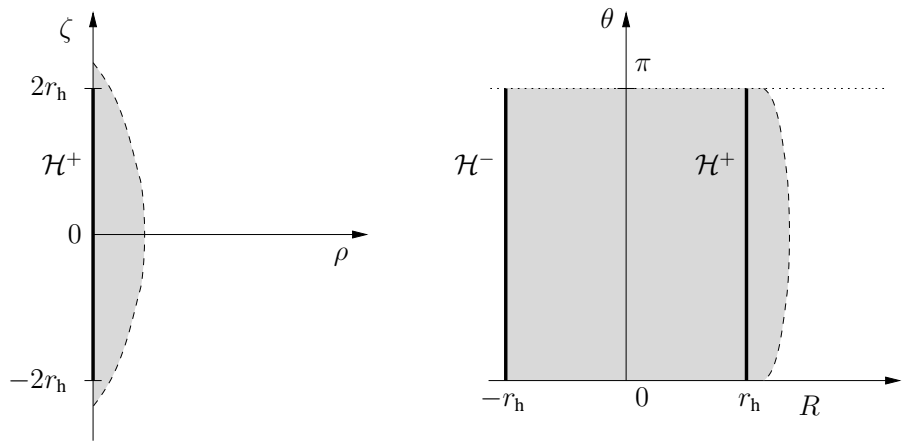}
 \caption{Sketch of a part of a black hole space-time
  in Weyl coordinates
 (\emph{left panel}) and Boyer-Lindquist type coordinates
 (\emph{right panel}). We
 study a vacuum region in an exterior vicinity of the event horizon and
 the interior of the black hole (grey areas).}
 \label{figure1}
\end{figure}

In a vacuum vicinity of the black hole's event horizon\footnote{For a
stationary black hole space-time, the immediate vicinity of the event
horizon must be vacuum, see. e.g. \cite{Bardeen}.} 
the Ernst potential is most easily introduced by utilizing the line
element in Weyl coordinates $(\rW,\zW,\varphi,t)$: 
\begin{equation}\label{WeylLineElement}
 \dd s^2 = \ee^{-2U}\left[\ee^{2k}(\dd\rW^2+\dd\zW^2)
            +\rW^2\dd\varphi^2\right]
           -\ee^{2U}(\dd t+a\dd\varphi)^2,
\end{equation}
where the metric potentials $U$, $k$ and $a$ are functions of $\rW$ and
$\zW$ alone.  
Along the rotation axis, $\rW=0, |\zW|\geq 2\rh$, the axial Killing
vector $\eta$ vanishes identically.  
The event horizon $\mathcal H^+$ is a degenerate surface when considered
in Weyl coordinates. It is located at $\rW=0, -2\rh\leq\zW\leq 2\rh$,
see Fig.~\ref{figure1}, left panel.  

The constant $\rh>0$ describes a coordinate radius of the event horizon
in Boyer-Lindquist type coordinates $(R,\theta,\varphi,t)$ which are
introduced via 
\beq\label{BoyerLcoord}
   \rW^2 = 4(R^2-\rh^2)\sin^2\theta,\qquad
   \zW   = 2R\cos\theta.
\eeq
These coordinates allow us to expand the metric
coefficients into the interior of the black hole\footnote{The interior ($-\rh< R < \rh$ ) 
corresponds to negative values of $\rW^2$.}. The horizons $\mathcal
H^\pm$ are located at $R=\pm\,\rh$, see Fig.~\ref{figure1}, right
panel. 

The complex Ernst potential $f$ combines metric functions,
\begin{equation}\label{ErnstPot}
 f=\ee^{2U}+\ii b,
\end{equation}
where the \emph{twist potential} $b$ is related to the coefficient $a$ via
\begin{equation}\label{a}
 a_{,\rW} = \rW\,\ee^{-4U} b_{,\zW},\qquad
 a_{,\zW} =-\rW\,\ee^{-4U} b_{,\rW},
\end{equation}
or, in terms of $R$ and $\theta$,
\begin{equation}\label{a2}
 a_{,R} = -2\sin\theta\,\ee^{-4U}b_{,\theta},\qquad
 a_{,\theta} = 2(R^2-\rh^2)\sin\theta\,\ee^{-4U}b_{,R}. 
\end{equation}
The vacuum Einstein equations are equivalent to the \emph{Ernst
equation} \cite {Ernst}, which reads in Weyl coordinates as
\begin{equation}\label{Ernst}
 (\Re f)\left(f_{,\rW \rW} + f_{,\zW \zW} + \frac{1}{\rW}f_{,\rW}\right)
 = f_{,\rW}^2 + f_{,\zW}^2 
\end{equation}
and in Boyer-Lindquist type coordinates:
\begin{equation}\label{Ernst1}
 (\Re f)\left[(R^2-\rh^2)f_{,RR}+2Rf_{,R}+f_{,\theta\theta}
               +\cot\theta f_{,\theta}\right]
  = (R^2-\rh^2)f_{,R}^2+f_{,\theta}^2.
\end{equation}
Note that $k$ can be calculated by a line integral
once $f$ is known, see e.g. \cite{AKMN2002}.

For convenience we introduce the following metric functions that are,
for a regular black hole,
positive and 
analytic in terms of $R$ and $\cos\theta$ in the black hole
vicinity, see \cite{Bardeen, Hennig2}: 
\begin{equation}\label{func1}
 \hat\mu:=4\ee^{2k-2U}(R^2-\rh^2\cos^2\!\theta),\qquad
 \hat u:=4(R^2-\rh^2)\ee^{-2U}-\frac{a^2}{\sin^2\!\theta}\ee^{2U}.
\end{equation}
At $\mathcal H^+$, the gravito-magnetic potential $\omega$, likewise
analytic in $R$ and $\cos\theta$ and defined through 
\begin{equation}\label{func2}
 \omega:=\frac{a\ee^{4U}}{4(R^2-\rh^2)\sin^2\!\theta-a^2\ee^{4U}},
\end{equation}
assumes the constant value $\omega^+$ describing the angular velocity of
the event horizon.

Because of the degeneracy of ${\cal H}^+$ in Weyl coordinates, the
potential $f$ is, for $\varrho=0$, only a $C^0$-function in terms
of $\zeta$. However, as the functions $\hat\mu$,
$\hat u$ and $\omega$, also $f$ is analytic with respect to the
Boyer-Lindquist type coordinates $R$ and $\cos\theta$. 

In the following sections we shall see that our conclusions only work if
(i) $\omega^+\neq 0$ and  
(ii) $b(\varrho=0, \zeta=2\rh)\neq b(\varrho=0, \zeta=-2\rh)$. However,
both situations may occur. If a rotating black hole with $J\neq 0$ is
dragged along  by the motion of a surrounding, sufficiently relativistic
counter-rotating torus then (i) the horizon angular velocity or (ii) the
black hole's Komar mass may vanish (see \cite{AnsorgPetroff2006}) which
would correspond to the two situations in question. Nevertheless, in
such a case our considerations can still be applied if one uses the
Ernst formulation in a {\em rotating} frame of reference
$\varphi'=\varphi+\Omega t$, $\Omega=\textrm{constant}$. It can be shown
that any such rotating system with $\Omega\neq \pm\kappa^+ A^+/(8\pi J)$
(`$\pm$' for the cases (i) and (ii) respectively) and $\Omega\neq 0$
could then be taken, where $\kappa^+$ is the black hole's surface
gravity of the event horizon. Therefore, without loss of generality we
shall henceforth assume that $\omega^+\neq 0$ and $b(\varrho=0,
\zeta=2\rh)\neq b(\varrho=0, \zeta=-2\rh)$. 
 
%%%%%%%%%%%%%%%%%%%%%%%%%%%%%%%%%%%%%%%%%%%%%%%%%%%%%%%%%%%%%%%%%%%%%%%%%
\section{B\"acklund transformation}\label{Baecklund1}

The B\"acklund transformation is a particular soliton method, which
creates a new solution from a previously known one. For the Ernst
equation this technique can be applied to construct a large number of
axisymmetric and stationary space-time metrics \cite{Neugebauer,Harrison, Kramer,
Ansorg2001,AKMN2002}. In this paper we 
consider the B\"acklund transformation in order to write an arbitrary
regular axisymmetric, stationary black hole solution $f$ in terms of a
potential $f_0$, which describes a space-time without a black hole, but
with a completely regular central vacuum region.

{\Thm\label{thm1}
Consider a regular axisymmetric and stationary black hole solution $f$
describing a sufficiently small exterior vacuum vicinity $V$
of the event horizon $\mathcal
H^+$. Then an Ernst potential $f_0=\ee^{2U_0}+\ii b_0$
of a space-time without a
black hole can be identified with the following properties: 
\begin{enumerate}
 \item $f_0$ is defined in a vicinity of the axis section
       $\rW=0, |\zeta|\leq 2\rh$. 
 \item In this vicinity, $f_0$ is an analytic function of $\rW$
       and $\zeta$ and an even function of $\varrho$.
 \item The axis values of $f_0$ in terms of those of $f$ for $\rW=0,
       |\zeta|\leq 2\rh$ are given by  
       \begin{equation}\label{f0}
        f_0=\frac{\ii\left[2\rh(b\pN+b\pS)-(b\pN-b\pS)\zW\right]f+4\rh
        b\pN b\pS} 
                 {4\rh f-\ii\left[2\rh(b\pN+b\pS)+(b\pN-b\pS)\zW\right]},
       \end{equation}
       where $b\pN=b(\varrho=0, \zeta=2\rh)$ and $b\pS=b(\varrho=0,
       \zeta=-2\rh)$ 
       (twist potential values at north and south pole of $\mathcal H^+$).
\end{enumerate}
From this Ernst potential $f_0$ the original potential $f$ can be
recovered in all of $V$ by means of an appropriate
B\"acklund transformation of the following form: 
\begin{equation}\label{Baecklund}
 f = \frac{\left|
   \begin{array}{ccc}
    f_0      & 1                 & 1\\
    \bar f_0 & \alpha_1\lambda_1 & \alpha_2\lambda_2\\
    f_0      & \lambda_1^2       & \lambda_2^2
   \end{array}
   \right|}
   {\left|
   \begin{array}{ccc}
    1      & 1                 & 1\\
    -1     & \alpha_1\lambda_1 & \alpha_2\lambda_2\\
    1      & \lambda_1^2       & \lambda_2^2
   \end{array}
   \right|},
\end{equation}
where
\begin{equation}\label{lambda}
 \lambda_i=\sqrt{\frac{K_i-\ii\bar z}{K_i+\ii z}},\qquad
 i=1,2,\qquad  K_1=-2\rh,\qquad  K_2=2\rh
\end{equation}
with the complex coordinates $z=\rW+\ii\zW$, $\bar z=\rW-\ii\zW$,
and $\alpha_1$, $\alpha_2$ are solutions to the Riccati equations
\begin{equation}\label{R1}
 \alpha_{i,z}=-(\lambda_i\alpha_i^2+\alpha_i)\frac{f_{0,z}}{2\ee^{2U_0}}
             +(\alpha_i+\lambda_i)\frac{\bar f_{0,z}}{2\ee^{2U_0}},
\end{equation}
\begin{equation}\label{R2}
 \alpha_{i,\bar z}=-\left(\frac{1}{\lambda_i}\alpha_i^2
                     +\alpha_i\right)\frac{f_{0,\bar z}}{2\ee^{2U_0}}
             +\left(\alpha_i+\frac{1}{\lambda_i}\right)
              \frac{\bar f_{0,\bar z}}{2\ee^{2U_0}}
\end{equation}
with
\begin{equation}\label{norm}
 \alpha_i\bar\alpha_i=1. 
\end{equation}
}

{\Pf
At first we show that the axis values of $f_0$ as given in \eref{f0}
form an analytic function with respect to $\zW$.  
Using the analyticity of $\hat{u}$ (being strictly positive) and
$\omega$ with respect to $\cos\theta$ as well as  Eqs.~\eref{func1} and
\eref{func2}, we may express $\ee^{2U}$ and $b$ on $\mathcal H^+$ as 
\begin{equation}\fl
 \ee^{2U}=-(\omega^+)^2\hat u\sin^2\!\theta,\qquad
 b =
 \frac{1}{2}\left[b\pN+b\pS+(b\pN-b\pS)\cos\theta\right]+A\sin^2\!\theta,
\end{equation}
with $A$ also being an analytic function in $\cos\theta$. Then \eref{f0} leads to
\begin{equation}
 f_0=\frac{(b\pN-b\pS)^2}{4[(\omega^+)^2\hat u-\ii A]}
     +\frac{\ii}{2}\left[b\pN+b\pS-(b\pN-b\pS)\cos\theta\right],
\end{equation}
which is analytic in $\cos\theta$ (and hence in $\zW$) and has a strictly positive
real part (recall that, without loss of generality, $\omega^+\neq 0$ and $b\pN\neq b\pS$, see end of Sec.~2).
Starting from the boundary values \eref{f0}, we can expand $f_0$ analytically and uniquely 
into some neighborhood of the axis part $\rW=0$, $|\zW|\le 2\rh$ by
virtue of the Ernst equation, see \cite{Hauser, Sibgatullin}.
This means that in contrast to $f$ the Ernst potential $f_0$ is analytic
with respect to the Weyl coordinates $(\rW,\zW)$ within this
neighborhood. Most important for later use, this axisymmetric expansion
of $f_0$ is {\em even} in $\rW$. 

%(This can easily be seen by explicitly constructing $f_0$. For
%this we refer to \cite{Hennig3}.)

Now we show that a B\"acklund transformation, applied to $f_0$,
returns our original Ernst potential $f$. To this end we need to choose appropriate
integration constants for the above Riccati equations. 

The equations \eref{R1} and \eref{R2} can be solved explicitly
on the horizon $\mathcal H^+$,
where $\lambda_i=\pm 1$ holds. We are free to choose the sign
convention $\lambda_1=-1$, $\lambda_2=1$.  
Then on $\mathcal H^+$ the equations \eref{R1} and \eref{R2} reduce to
\begin{equation}
 \alpha_{i,\zW}=\frac{\alpha_i+\lambda_i}{2\ee^{2U_0}}
               \left(-\lambda_i\alpha_i f_{0,\zW}+\bar f_{0,\zW}\right)
\end{equation}
with the solution
\begin{equation}\label{alpha}
 \alpha_1(\zW) = -\frac{\bar f_0(\zeta)+\ii\gamma_1}
                       {f_0(\zeta)-\ii\gamma_1},\qquad
 \alpha_2(\zW) = \frac{\bar f_0(\zeta)+\ii\gamma_2}
                       {f_0(\zeta)-\ii\gamma_2}.
\end{equation}
The integration constants $\gamma_i$ are
\emph{real} numbers in order to guarantee \eref{norm}. 

In terms of the coordinates $R$ and $\theta$, the above B\"acklund transformation Eq.~\eref{Baecklund} reads generally as follows:
\begin{equation}\label{funcf}
 f=
        \frac{[\alpha_1(R+\rh\cos\theta)-\alpha_2(R-\rh\cos\theta)]f_0
         +2\rh\bar f_0}
        {\alpha_1(R+\rh\cos\theta)-\alpha_2(R-\rh\cos\theta)-2\rh}.
\end{equation}
On the horizon $\mathcal H^+$ ($R=\rh$), Eqns. \eref{alpha} and
\eref{funcf} lead to an Ernst potential with the values $\ii\gamma_1$
and $\ii\gamma_2$ on the north and south pole of $\mathcal H^+$
respectively.  
Now, if we choose consistently $\gamma_1=b\pN$ and $\gamma_2=b\pS$, then
\eref{funcf} becomes equivalent to \eref{f0}. 
Since in a vicinity of $\mathcal H^+$ the Ernst
potential $f$ is uniquely determined by its horizon values
(due to a theorem by Hauser and Ernst \cite{Hauser}),
we recover the original solution $f$ in this vicinity.
~    \qquad\hfill $\Box$
}
%%%%%%%%%%%%%%%%%%%%%%%%%%%%%%%%%%%%%%%%%%%%%%%%%%%%%%%%%%%%%%%%%%%%%%%%%
\section{The Ernst potential on the Cauchy horizon}

In this section we expand the exterior Ernst potential $f$ into the
interior of the black hole, i.e.~to the region $R\in[-\rh,\rh]$. As
mentioned in Sec.~2, for
regular black holes the Ernst potential $f$ is analytic with respect to
$R$ and $\cos\theta$ in an exterior vicinity of ${\cal H}^+$. Hence we
can expand it analytically into an \emph{interior} vicinity of $\mathcal
H^+$.  
Then, due to a theorem by Chru\'sciel (theorem 6.3 in
\cite{Chrusciel}\footnote{We note that the interior space-time region of
axisymmetric and stationary black holes is closely related to Gowdy
space-times. In particular, we obtain Chru\'sciel's form of the Gowdy
space-time metric
by substituting $R=\rh\cos T$ and $\theta=\psi$. More
information will be presented in \cite{Hennig3}.}), 
the potential $f$ exists as a regular solution of the interior Ernst
equation for all values $(R,\cos\theta)\in(-\rh,\rh]\times[-1,1]$, 
i.e.~within a region that only excludes the Cauchy horizon $\mathcal
H^-$ ($R=-\rh$).  
In the following we obtain an explicit formula for
$f$ on $\mathcal H^-$ in terms of the boundary data on $\mathcal H^+$,
which shows that $f$ is also regular on $\mathcal H^-$ provided that
$J\neq0$ holds.

\begin{figure}
 \centering
 \includegraphics[scale=0.9]{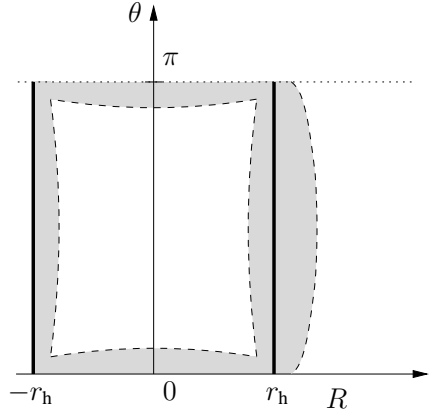}
 \caption{The seed function $f_0$ can be regularly defined
 in at least the grey
 areas.}
 \label{figure2}
\end{figure}

A crucial role for our considerations is played  by the fact that $f_0$ is even
in $\varrho$, see discussion in Sec.~3. 
Hence, in terms of the Boyer-Lindquist type coordinates
\eref{BoyerLcoord}, $f_0$ is an analytic function of
$(R^2-\rh^2)\sin^2\!\theta$ and $R\cos\theta$. The analytic expansion of
$f_0$ into the region $R<\rh$ retains this property. As a consequence we
find that the interior boundary values $f_0(R,\cos\theta=\pm1)$,
$ -\rh\leq R\leq \rh$, as well as $f_0(-\rh,\cos\theta)$ are given in
terms of the values at $R=\rh$. Also it follows that $f_0$ is
regularly defined in a sufficiently small vicinity of the boundary of
the interior region, see \fref{figure2}. Specifically we obtain
\beq
	f_0(R=\,-\,\rh\,,\cos\theta) = f_0(R=\,+\,\rh\,,-\cos\theta).
\eeq
These properties allow us to construct $f$ on $\mathcal H^-$ from $f_0$ via the B\"acklund
transformation \eref{Baecklund}. As done for the proof of theorem \ref{thm1}, we
solve the Riccati equations \eref{R1} and \eref{R2}, now on $\mathcal H^-$ and for $\cos\theta=\pm 1$ 
in accordance with the solution on $\mathcal H^+$,
thereby obtaining a unique continuous function $f$ defined on the entire boundary of the interior region.
In particular we find the following.

{\Thm Any Ernst potential $f$ of a regular axisymmetric and stationary black hole
 space-time with angular momentum $J\neq 0$
 can be regularly extended into the interior of the black hole up to and including an interior Cauchy horizon, described by $R=-\rh$ in Boyer-Lindquist type coordinates $(R,\theta)$. The values of $f$ on the Cauchy horizon are given by
\begin{equation}\label{fCH}
 \fl f(R=-\rh,\cos\theta)=\frac{\ii\left[\delta_1+\delta_2
                  -(\delta_1-\delta_2)\cos\theta
                \right]f_0(R=\rh,-\cos\theta)+2\delta_1\delta_2}
                {2f_0(R=\rh,-\cos\theta)
                -\ii\left[\delta_1+\delta_2+(\delta_1-\delta_2)\cos\theta
                \right]}
\end{equation}
with\footnote{As we shall see in \eref{J2}, $\delta_1$ and $\delta_2$ are
well-defined because $b\pN-b\pS+2(b_{,\theta\theta})\pN\neq 0$ for $J\neq 0$. 
For $J\to0$ we have $|\delta_{1/2}|\to\infty$ and $f|_{\mathcal H^-}$ diverges.}
\begin{equation}
 \delta_1 = \frac{b\pS(b\pN-b\pS)+2b\pN(b_{,\theta\theta})\pN}
                 {b\pN-b\pS+2(b_{,\theta\theta})\pN},\qquad
 \delta_2 = \frac{b\pN(b\pN-b\pS)+2b\pS(b_{,\theta\theta})\pN}
                 {b\pN-b\pS+2(b_{,\theta\theta})\pN}          
\end{equation}
where the scripts {\rm `+'} and {\rm `N/S'} indicate that the corresponding value of $b$ or its second $\theta$-derivative has to be taken at the event horizon's north or south pole respectively. The values of the seed solution $f_0$ for $R=\rh$ follow via \eref{f0} from $f$ on the event horizon. For $J\to0$ the Cauchy horizon becomes singular.}

%%%%%%%%%%%%%%%%%%%%%%%%%%%%%%%%%%%%%%%%%%%%%%%%%%%%%%%%%%%%%%%%%%%%%%%%%
\section{A universal equality}

With the relation between $f|_{\mathcal H^+}$ and $f|_{\mathcal H^-}$ we
are able to prove the equality 
(\ref{UnivEquality}). Angular momentum $J$ and horizon areas $A^\pm$ are
given as follows in terms of the Ernst potential: 
\bea\label{J}
  \fl J  &=&  \frac{1}{8\pi}\oint\limits_\mathcal{H^\pm}\eta^{a;b}\dd S_{ab}
     = -\frac{1}{16}\int\limits_0^\pi\hat u^2\omega_{,R}\big|_\mathcal{H^\pm}
       \sin^3\!\theta\,\dd\theta
     = 2\frac{b\pS-b\pN}{(b_{,R}^{\ 2})\pN}-\frac{\rh}{4}(b_{,R})\pN,\\
  \fl A^\pm &=& 2\pi\int\limits_0^\pi\sqrt{\hat\mu\hat u}\big|_\mathcal{H^\pm}
     \sin\theta\,\dd\theta
   = 4\pi \hat u\pmN=\pm\frac{32\pi\rh}{({\ee^{2U}}_{,R})\pmN},
\eea
Here we used  \eref{func1}, \eref{func2} and equation (18) of \cite{Hennig2} (together with a corresponding version valid on  $\mathcal H^-$). The derivative $f_{,R}$ on $\mathcal H^\pm$ can be
calculated from $f$ and its $\theta$-derivatives by considering the Ernst equation \eref{Ernst1} which becomes degenerate
at $R=\pm\rh$.
%obtaining
%\begin{equation}
%  f_{,R} = \frac{1}{\pm 2\rh}
%    \left(\frac{f_{,\theta}^2}{\ee^{2U}}-f_{,\theta\theta}
%    -\cot\theta f_{,\theta}\right)\quad\textrm{on}\quad\mathcal H^\pm.
%\end{equation}
In particular, at the north and south poles where $\ee^{2U}=0$ and
$f_{,\theta}=0$, we obtain via L'Hospital's rule 
\begin{equation}\label{fNP}
 f_{,R}=\pm\ii\frac{f_{,\theta\theta}\,b_{,\theta\theta}}
         {\rh {\ee^{2U}}_{,\theta\theta}}\quad\textrm{for}\quad
         R=\pm\rh,\quad \sin\theta=0.
\end{equation}

With the formula \eref{fCH} for the Ernst potential on the Cauchy
horizon, we finally arrive after some calculation at
\begin{equation}
 A^+=-32\pi\rh^2\left.\frac{{\ee^{2U}}_{,\theta\theta}}
      {b_{,\theta\theta}^{\ 2}}
     \right|\pN,\qquad
 A^-=-8\pi\rh^2\frac{\left(b\pN-b\pS+2(b_{,\theta\theta})\pN\right)^2}
     {({\ee^{2U}}_{,\theta\theta}\,b_{,\theta\theta}^{\ 2})\pN},
\end{equation}
\begin{equation}\label{J2}
 J=-2\rh^2\frac{b\pN-b\pS+2(b_{,\theta\theta})\pN}
    {(b_{,\theta\theta}^{\ 2})\pN},\qquad
 \omega^+ = \frac{(b_{,\theta\theta})\pN}{4\rh}\neq 0. 
\end{equation}
Note that ${\ee^{2U}}_{,\theta\theta}< 0$ as $A^+>0$ for regular black holes. 
Together with the results in \cite{Hennig2} we thus find the following.
{\Thm 
 Every regular axisymmetric and stationary black hole with non-vanishing angular momentum $J$
 satisfies the relation $(8\pi J)^2=A^+ A^-$ where $A^\pm$ are the horizon areas of event ($\mathcal H^+$) and Cauchy horizon ($\mathcal H^-$).
 If in addition the black hole is sub-extremal (i.e. if there
 exist trapped surfaces in every sufficiently small interior vicinity of
 $\mathcal H^+$, see \cite{Booth}), then the following inequalities hold:
 $A^- < 8\pi|J| < A^+$. Moreover, sub-extremal black holes with $J\neq0$ have no trapped surfaces in 
sufficiently small interior vicinities of $\mathcal H^-$.
}

%%%%%%%%%%%%%%%%%%%%%%%%%%%%%%%%%%%%%%%%%%%%%%%%%%%%%%%%%%%%%%%%%%%%%%%%%
\ack
We would like to thank Vincent Moncrief and Alan Rendall for many
valuable discussions. 
This work was supported by the Deutsche Forschungsgemeinschaft (DFG)
through the Collaborative Research Centre SFB/TR7
\lq\lq Gravitational wave astronomy\rq\rq. 

%%%%%%%%%%%%%%%%%%%%%%%%%%%%%%%%%%%%%%%%%%%%%%%%%%%%%%%%%%%%%%%%%%%%%%%%%

%%%%%%%%%%%%%%%%%%%%%%%%%%%%%%%%%%%%%%%%%%%%%%%%%%%%%%%%%%%%%%%%%%%%%%%%%

\end{document}